**Tuning the properties of metal surfaces by alloying: a DFT study of $H_2$, $O_2$, and $H_2O$ adsorption on Ni-Fe surfaces**

Changyuan Li[a] and Sergio Tosoni*[b]

[a]Department of Material Science & Technology, Nanjing University of Aeronautics and Astronautics, Nanjing 211106, China.

[b]Dipartimento di Scienza dei Materiali, Università di Milano-Bicocca, via Roberto Cozzi 55, 20125 Milano (Italy). Email: sergio.tosoni@unimib.it

**Abstract**

The adsorption and dissociation of $H_2$, $O_2$, and $H_2O$ on Ni-Fe alloys with variable Fe:Ni ratio are studied by means of Density Functional Theory calculations. The alloy composition deeply influences the thermochemistry of the adsorption and dissociation processes, with relevant implications to catalysis and electrocatalysis. For large concentration of Fe, the adsorption of $H_2$ is facilitated. On the contrary, $O_2$ is bound more strongly on Ni(111). The dissociation of all three molecules is favourable on all considered supports. However, the $O_2$ and $H_2O$ dissociation is more favourable on NiFe alloys compared to Ni, while the dissociation of H2 has a similar thermodynamic profile on pure Ni and $Ni_2Fe$ alloy, and is less favourable on NiFe.

**Keywords**

DFT; NiFe alloys; Catalysis; Electrocatalysis;

## 1. Introduction

Many industrial chemical processes, from fine chemical synthesis to combustion control, rely on catalysts to improve their yield and sustainability. This is crucial in the actual frame, where finding efficient pathways toward the production of energy, fuels and chemicals from renewable sources is mandatory. In many cases, the catalytic devices actually in use fall in the domain of heterogeneous catalysis.[1] The role of transition metal surfaces in this domain is notoriously very important, sometimes also in nanostructured forms.[2,3] This, in turn, implies the challenge to find efficient materials based on non-noble and earth-abundant elements, such as Fe or Ni, for both catalytic and electrocatalytic applications.[4–6] Nickel, in particular, has driven a remarkable attention for his high activity in environmentally relevant reactions such as the water gas shift (WGS) and steam methane reforming (SMR). A fundamental aspect common to both WGS and SMR is the determining role played by water dissociation for the overall reaction's dynamics. Despite the very long history of fundamental surface science investigations on the behaviour of water on transition metal surfaces,[7] a thorough computational study of $H_2O$ adsorption and dissociation on the most stable nickel surfaces has highlighted the non-trivial role played by the surface morphology in promoting the dissociative adsorption of water, and binding the related fragments on the most stable nickel surfaces.[8]

Recently, the thermochemistry of all elementary steps relevant to the WGS reaction have been calculated on Ni(111),[9] a well-known active support for this reaction of paramount relevance in chemistry, also considering the effects of the various nickel surfaces.[10]

Since many years, the possibility to tune the chemical properties of metals, even profiting of synergistic effects, by alloying has been investigated.[11,12] In a broad sense, the mixing of metal atoms of type A in a matrix formed by B elements will both act on the collective, physical properties exhibited by the AB alloy, and on the local binding environment of A atoms surrounded by B atoms.[13] Many factors, thus, may play a role, from the A:B ratio, to the nanostructuring of the alloy and its homogeneous or segregated nature.[14] Focusing on the adsorption of molecular or atomic species on alloy surfaces, one can distinguish between ligand effects (arising from the modification of the electronic structure of the surface due to alloying) and ensemble effects (due to the local chemical composition of the adsorption site).[15] The intermixing of noble metals have led to alloys with interesting properties, as recently shown for Ag-Au[16] or Ag-Pd.[17] However, another relevant advantage of alloying is precisely the chance to spare a relevant fraction of precious raw metals and substitute it with an earth-abundant one. Alloying platinum with Ni, for instance, was recently shown to be a good way to yield reaction rates comparable to Pt(111) for water splitting and $O_2$, $H_2$ recombination, a promising result toward sparing precious metals.[18] Following a similar idea, the catalytic dissociation of water has been investigated over the (111) surface of Ni-Pt alloys, also considering the surface stability as a function of the chemical composition.[19] The electrocatalytic properties of several binary and ternary Ni-based alloys have been scanned in a computational study on the hydrogen oxidation reaction, a chemical process relevant to fuel cells, showing that Ti-Ni-Cu alloys could be the optimal substrates for this reaction. [20] The $CO_2$ dissociation process has been as well investigated computationally over a wide range of Ni-based alloys.[21] Interestingly, the CO binding energy resulted scarcely sensitive to the surface composition, while the O adsorption energy displayed a remarkable variation. The latter was attributed to the ensemble effect rather than to the ligand effect.

It must be, moreover, stated that, beside their activity, Ni-based alloys display a promising durability under operative electrochemical conditions.[22]

In particular, Ni-Fe alloys display very interesting features; low cost, abundance, and good activity in several electrocatalytic or catalytic processes such as oxygen reduction or hydrogen oxidation.[23–26] Iron is a metal which is both cheap, and with remarkable catalytic properties, as shown for the water splitting reaction by single iron atoms implanted in carbonaceous supports.[27] As for Ni-Fe alloys, the main reason to incorporate a certain amount of iron in a nickel matrix is to improve the too low hydrogen binding energy on Ni.[28] The different oxygen affinity of Ni and Fe, in turns, determine the structural properties of the reactive oxygen overlayer formed on Ni-Fe alloys under oxidative conditions, with important implications on the chemical reactivity.[25] The studies mentioned above are definitely indicating a general suitability of Ni:Fe alloys for catalytic applications. However, a proof of concept on the role of alloying in determining the catalytic properties is still lacking.

Aiming at a deeper rationalization of the structure-reactivity relationship in Ni-Fe alloys with respect to some of the key-reactants for catalytic or electrocatalytic applications, we hereby study the adsorption and dissociation of hydrogen, oxygen, and water on Ni(111) and on two Ni-Fe model structures, whose Fe:Ni ratio varies from 0.33 to 0.50, by means of Density Functional Theory (DFT) calculations.

2. **Computational details**

All calculations are done with the code VASP 5.[29,30] The Perdew, Burke and Hernzerhof (PBE) exchange-correlation is adopted.[31] The long-range dispersion forces are accounted for recurring to the D3 scheme from Grimme and the Becke-Johnson damping function.[32,33] All calculations are spin polarized. The plane-wave basis set is expanded within a kinetic energy cutoff of 400 eV. The core electrons are described with the Projector Augmented Wave (PAW) method.[34,35] Ni(3d,4s), Fe(3d,4s), H(1s) and O(2s,2p) are treated as valence electrons. The metallic surfaces are represented with a 4-layers slab model based on the Ni(111) structure, where a given number of Ni atoms in the topmost layer are substituted by Fe atoms. The two topmost layers are relaxed during the structural optimization process, while the bottom layer is kept fixed at the bulk geometry. A conjugated gradient algorithm is adopted for the structural relaxation. A lattice constant of 3.514 Å is calculated for bulk Ni. A 2√3×3 supercell is used to simulate the adsorption and dissociation of $H_2$, $O_2$, and $H_2O$. A layer of empty space as thick as 15 Å is included in the supercell to avoid spurious interactions with the replica of the slab. Dipole and quadrupole corrections are applied along the non-periodic direction. A mesh of 3×3×1 K-points is adopted to sample the reciprocal space. The adsorption free energy is defined as:

$\Delta G = G(molecule^*) - (G(molecule) + G(surface))$

where $G(molecule^*)$ is the free energy of a molecule adsorbed on the surface, while $G(molecule)$ and $G(surface)$ are the free energy of the gas-phase molecule and the clean surface, respectively.

$\Delta G$ at T= 298 K are obtained by including the Zero Point Energy (ZPE) and the measured standard entropy of the gas-phase molecular species.[36] The ZPE is evaluated by calculating the harmonic vibrational frequencies of the adsorbed molecules. The Hessian matrix of atomic displacements is constructed by moving the adsorbed atoms and the metal atoms directly bound to them.

## 3. Results and Discussion

### 3.1 *Surface properties of NiFe alloys*

We begin by comparing the main structural properties of Ni-Fe alloys, compared to pure Ni(111). Nickel crystallizes with a face-centred cubic structure (Fm-3m space group),[37] and its lowest-energy surface is the (111).[38] Its ground state is ferromagnetic, and with the presently adopted computational setup, a magnetic moment of 0.66 $\mu_B$ is attributed to each Ni atom (Table 1), in excellent agreement with experimental measurements of the 3d spin contribution to the magnetic moment of Ni.[39] The calculated work function, 5.01 eV, is very close to the experimental value (5.15 eV)[40] and to previous theoretical results obtained with a comparable computational methodology (5.08 eV).[41]

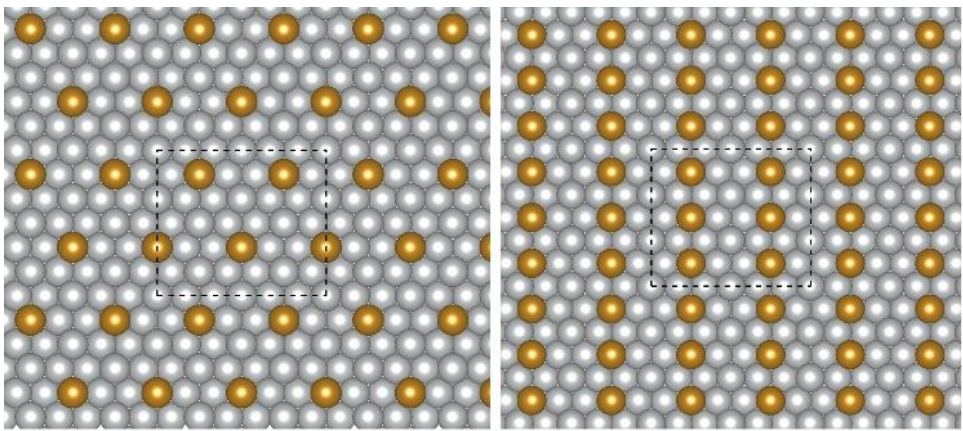

**Figure 1**. Top view of $Ni_2Fe/Ni(111)$, left, and NiFe/Ni(111), right. Ni atoms are displayed in grey, Fe atoms in orange. The computational supercell is indicated with the black dashed line.

**Table 1**. Work function (f, eV), mean atomic magnetization (<M>, $\mu_B$), mean surface rumpling (<R>, Å).

|  | f (eV) | (<M> ($\mu_B$) | <R> (Å) |
|---|---|---|---|
| Ni(111) | 5.01 | 0.66 | 0.00 |
| Ni$_2$Fe/Ni(111) | 4.98 | Ni=:0.56<br><br>Fe= 2.99 | 0.05 |
| NiFe/Ni(111) | 4.98 | Ni=:0.61<br><br>Fe= 2.84 | 0.05 |

The models of Ni$_2$Fe and NiFe alloys are generated simply by substituting one third, or one half, of the surface Ni atoms with Fe atoms, Fig. 1. Such a model, in the case of Ni$_2$Fe, was previously adopted to study the activation and decomposition of CH$_4$ and CO$_2$.[26] The rationale to adopt this kind of model relies on the fact that nickel-iron composites where Ni is the dominant element tends to preserve a crystal structure close to cubic fcc Ni.[25] It is thus possible to simulate the surfaces of NiFe alloys with variable Fe content by keeping the simple and well-defined structure of bulk Ni as a support. In this work, we compare the different adsorptive properties of bare Ni(111) and two alloys with Ni:Fe ratios of 1.5 and 1, respectively, toward three simple molecules of paramount relevance in catalytic and electrocatalytic properties: hydrogen, oxygen, and water.

A first, interesting finding is reported in Table 1. While the work function of the NiFe alloys is not affected by the iron concentration at the surface, the magnetic moments of the surface atoms change remarkably: for nickel, we observe a small decrease of the magnetization from 0.66 $\mu_B$ in bare Ni(111) to 0.56 $\mu_B$ in Ni$_2$Fe/Ni(111) and 0.61 $\mu_B$ in NiFe/Ni(111). The magnetic moment of iron atoms is 2.99 $\mu_B$ in Ni$_2$Fe/Ni(111) and 2.84 m$_B$ in NiFe/Ni(111). These values are rather high compared to what is calculated for pure, bulk iron (2.2 $\mu_B$),[42] where the hybridization of the atomic orbitals forming the bands causes a decrease in magnetization from the value of an isolated Fe atom in 4s$^2$, 3d$^6$ electron configuration, i.e. 4. Notably, iron clusters up to several hundreds of atoms display larger atomic magnetization with respect to iron bulk, with interesting implications for their physical and chemical properties.[42] In our Ni-Fe models, the large magnetic moment of the iron atoms should be attributed to hybridization effects, rather than to an actual charge transfer between Ni and Fe, which is not observed by analysing the atomic populations. Nevertheless, this feature may be of interest for the application of Ni-Fe alloys in magnetism or electronics. An aspect of more specific relevance for the surface chemistry is related to the small protrusion (0.05 Å) exhibited by the Fe atoms hosted on surface Ni lattice sites, despite their comparable atomic radii. This fact may be relevant when atoms or small molecules are adsorbed on the surface (vide infra).

### 3.2 *Adsorption of $H_2$, $O_2$, and $H_2O$ on Ni and NiFe alloys*

We now discuss the adsorption of three simple probe molecules ($H_2$, $O_2$ and $H_2O$) on Ni(111), $Ni_2Fe$/Ni(111) and NiFe/Ni(111). The Ni(111) surface is topologically quite simple, and displays Ni-top, bridge, or hollow adsorption sites. The landscape is slightly more complicated in the case of Ni-Fe alloys, featuring also Fe-top, $Ni_2Fe$ and $NiFe_2$ hollows. In this work, we have explored all possible adsorption sites for the three molecules (see ESI), and report the most stable adsorption configuration in Fig. 2.

The corresponding adsorption free energies are reported in Fig. 3. For all three molecules, we consider a physisorption step, where the adsorbed molecule remains intact, followed by a dissociation, where molecular fragments remain chemisorbed at the metal surface, namely:

$(H_2)_{gas} \rightleftarrows (H_2)^*$ 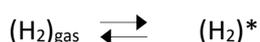

$(H_2)^* \rightleftarrows (H)^* + (H)^*$ 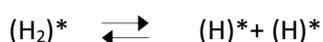

$(O_2)_{gas} \rightleftarrows (O_2)^*$ 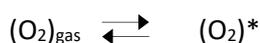

$(O_2)^* \rightleftarrows (O)^* + (O)^*$ 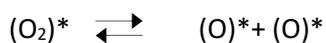

$(H_2O)_{gas} \rightleftarrows (H_2O)^*$ 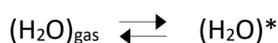

$(H_2O)^* \rightleftarrows (OH)^* + (H)^*$ 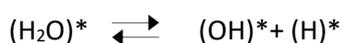

The reactivity of the surface is here evaluated based only on the free energy change associated to each elementary step.[43] We notice that $H_2$ is weakly physisorbed in molecular form on Ni(111), similarly to what previously reported,[9] and $Ni_2Fe$/Ni(111), with an H atom pointing toward a hollow site. The distance from the surface is, accordingly, in the range for weak physisorption (around 3 Å). The adsorption free energy is positive by 0.32 eV and 0.34 eV, due to entropic factors. The situation is different on NiFe/Ni(111), where a di-hydrogen complex on top of a Fe atom is formed, with Fe-H distances of 1.67 Å. The thermodynamic barrier to form $H_2^*$ is remarkably smaller (+0.14 eV) in this case. The iron atoms exposed at the NiFe/Ni(111) permit thus a remarkable stabilization of the physisorbed intermediate. This is a first example of relevant ligand effect, given the different behaviour observed for NiFe/Ni(111) compared to $Ni_2Fe$/Ni(111) and pure Ni(111).

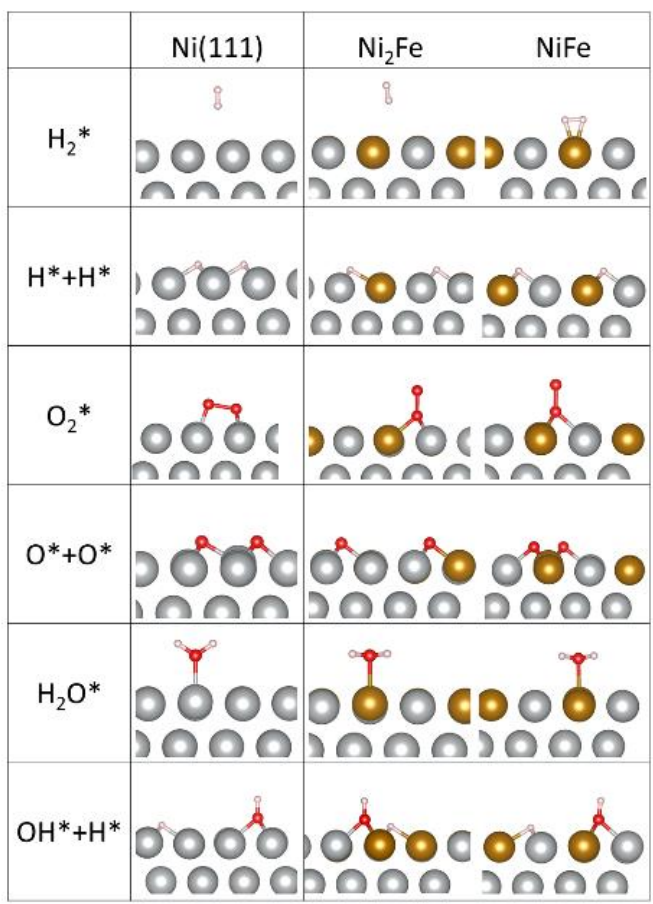

**Figure 2**. Side views of most stable adsorbed species on Ni(111), Ni$_2$Fe/Ni(111), and NiFe/Ni(111).

The dissociation of the hydrogen molecule on the metallic surface leads to the formation of two separated hydride species, Fig. 2, preferentially bound to Ni- or Ni-Fe hollow sites. With respect to H$_2$*, we observe a stabilization on all supports, Fig. 3a, and the (H*+H*) adducts have similar free energy of adsorption, -0.67 eV on Ni(111), -0.66 eV on Ni$_2$Fe/Ni(111), and -0.60 eV on NiFe/Ni(111). It is worth noting that the opposite process, i.e. the formation and subsequent evolution of a H$_2$ molecule would be remarkably more favourable on NiFe compared to Ni$_2$Fe or Ni, facing a barrier of 0.74 eV, compared to almost 1 eV for Ni$_2$Fe/Ni(111) and Ni(111).

The oxygen molecule displays a different preferential orientation on nickel and on Ni-Fe alloys. On bare Ni(111), the oxygen molecule lies almost horizontally, with a small tilt angle with respect to the metal surface. One O atom is adsorbed on a Ni-top site, while the other is di-coordinated at a bridge site. The O-O interatomic distance is elongated by 0.23 Å with respect to the gas phase molecule. On the Ni-Fe alloys, O$_2$* is preferentially adsorbed in a vertical arrangement, with an O atom pointing to a Ni$_2$Fe or NiFe$_2$ hollow site on Ni$_2$Fe/Ni(111) and NiFe/Ni(111), respectively. The adsorption ΔG at room conditions, Fig. 3b, shows that the adsorption of O$_2$* is almost thermoneutral on Ni$_2$Fe/Ni(111), -0.13 eV, slightly exergonic on NiFe/Ni(111),

-0.29 eV, and remarkably exergonic on Ni(111) (-1.34 eV). Notably, also the adsorption geometry of $O_2$*
changes from almost flat in top position (pure Ni) to vertical in hollow sites (both Ni-Fe alloys).

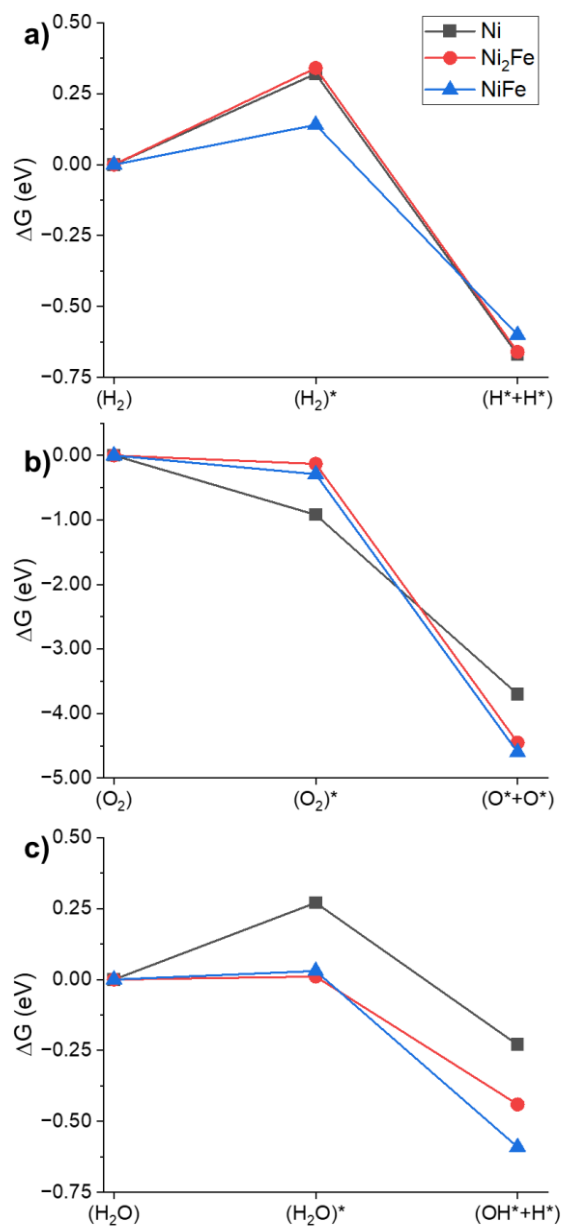

**Figure 3**. ΔG of adsorption and dissociation of a) $H_2$, b) $O_2$, and c) $H_2O$ on Ni(111), $Ni_2Fe$/Ni(111), and NiFe/Ni(111).

A strong gain of free energy is envisaged on all supports passing from molecularly adsorbed $O_2$* to dissociatively chemisorbed O* atoms on the surface, Fig. 3b. This gain is larger on Ni-Fe alloys, where the (O*+O*) adducts are stabilized by 4.45 eV ($Ni_2Fe$/Ni(111)) and 4.60 eV (NiFe/Ni(111)) with respect to the gas-phase $O_2$ and the clean surface. A smaller free energy, -3.70 eV, is reported for Ni(111). The O* atoms are three-fold coordinated on hollow sites both on Ni(111) and on Ni-Fe alloys, Fig. 2. On the latter, however, the hollow sites expose both Ni and Fe atoms, which interact strongly with O*.

For oxygen-related species on NiFe/Ni(111), two different minima were found, where either $O_2$* or O* is bound on a $Ni_2Fe$ hollow site, or on a $NiFe_2$ one, see Table S1 in ESI. An increase in adsorption energy as large as 15-20% is reported for $NiFe_2$ hollows compared to $Ni_2Fe$: this relevant ensemble effect further shows that the most oxophylic character of Fe compared to Ni dictates the stabilization of oxygen species on the surface.

The water molecule, finally, interacts preferentially with its oxygen molecule on top of a metal atom, Fig. 2. On bare Ni(111), it preserves an almost vertical orientation, while on Ni-Fe alloys it is remarkably tilted, fig. 2. Energetically, the adsorption of a water molecule has a negligible ∆G on $Ni_2Fe$/Ni(111) and NiFe/Ni(111), Fig. 3c, while a thermodynamic barrier of 0.27 eV must be overcome on Ni(111). The adsorption energy for pure nickel (calculated neglecting ZPE and entropic contributions, as in Table S1, ESI) compares well with previous results.[8] The water molecule can be split at the metal surface in a hydroxyl (OH*) and a hydride (H*) species, Fig. 2. The OH* species is adsorbed vertically, with the O atom bound to a hollow site, while the H* atom, as described above, bound to a three-fold hollow site. On Ni(111), the gain of free energy is moderated (∆G = -0.23 eV). The Ni-Fe alloys show a stronger activity toward water dissociation (-0.44 eV for $Ni_2Fe$/Ni(111) and -0.59 eV for NiFe/Ni(111)). The splitting of the water molecule is thus remarkably more favourable on alloys (NiFe in particular) compared to pure nickel. This is mostly related to the increased capability of binding O* obtained by doping Ni surfaces with Fe.

4. **Conclusions**

Alloying with iron is an effective strategy to tune the adsorptive properties of nickel. In particular, NiFe/Ni(111) has both a smaller barrier to bind $H_2$, and a smaller barrier to recombine H*, compare to bare Ni or Ni-Fe with lower iron content. Ni-Fe alloys, on the contrary, have a smaller affinity than Ni for molecular $O_2$, but tend to bind more strongly O*. The adsorption of water, and the subsequent stabilization of OH and H species on the surface, is remarkably enhanced in presence of iron species at the Ni surface; overall, this facilitates the splitting of the water molecule on iron-containing surfaces with respect to pure nickel. Ligand effects (related to the different chemical nature of the surface) matter in determining the surface properties of the Ni-Fe alloys, but ensemble effects (related to the different chemical composition of the hollow binding sites) can also be relevant.

The larger magnetic moment of the Fe atom in the alloys, with respect to pure Fe, is also worth to be noted for potential technological applications.

**Acknowledgements**

Sergio Tosoni acknowledges financial support from ICSC – Centro Nazionale di Ricerca in High Performance Computing, Big Data and Quantum Computing, funded by European Union – NextGenerationEU and from the Italian Government (PRIN Project 2022LS74H2-SUMCAR). Changyuan Li acknowledges financial support from CSC – China Scholarship Council in staying and study support (File No.202206830150).

**Authors Contribution**

Changyuan Li: investigation and data management. Sergio Tosoni: conceptualization and supervision. Both authors equally contributed to writing.

**Conflicts of interest**

The authors declare no conflicting interests.

# Supporting Information:

# Tuning the properties of metal surfaces by alloying: a DFT study of $H_2$, $O_2$, and $H_2O$ adsorption on Ni-Fe surfaces

Changyuan Li[a] and Sergio Tosoni*[b]

[a.] Department of Material Science & Technology, Nanjing University of Aeronautics and Astronautics, Nanjing 211106, China.
[b.] Dipartimento di Scienza dei Materiali, Università di Milano-Bicocca, via Roberto Cozzi 55, 20125 Milano (Italy). Email: sergio.tosoni@unimib.it

**Table S1**: Adsorption energy (eV) on various surface adsorption sites

| Surface | Molecule | Ni-Top | Fe-Top | Hollow ($Ni_3$) | Hollow ($Ni_2Fe$) | Hollow ($NiFe_2$) |
|---|---|---|---|---|---|---|
| Ni | $H_2$* | -0.10 | | -0.11 | | |
| $Ni_2Fe$ | | -0.09 | -0.08 | | -0.09 | |
| NiFe | | -0.08 | -0.36 | | -0.09 | -0.09 |
| Ni | | -1.33 | | -1.30 | | |
| $Ni_2Fe$ | | -1.34 | -1.34 | | -1.34 | |
| NiFe | | -1.26 | -1.29 | | -1.31 | -1.27 |
| Ni | $O_2$* | -1.66 | | -0.77 | | |
| $Ni_2Fe$ | | | | | -0.93 | |
| NiFe | | | | | -0.85 | -1.04 |
| Ni | O* + O* | -1.66 | | -4.31 | | |
| $Ni_2Fe$ | | -4.99 | -5.03 | | -5.03 | |
| NiFe | | -5.08 | -5.16 | | -4.69 | -5.44 |
| Ni | $H_2O$* | -0.38 | | -0.26 | | |
| $Ni_2Fe$ | | -0.41 | -0.66 | | -0.45 | |
| NiFe | | -0.38 | -0.64 | | -0.64 | -0.48 |
| Ni | OH* + H* | -0.56 | | -0.64 | | |
| $Ni_2Fe$ | | -0.65 | -0.52 | | -0.70 | |
| NiFe | | -0.85 | -0.68 | | -0.68 | -0.83 |